\documentclass[
 aip,
 amsmath,amssymb,
 reprint,
]{revtex4-1}

\usepackage{graphicx}
\usepackage{dcolumn}
\usepackage{booktabs}
\usepackage{multirow}
\usepackage{siunitx}
\usepackage{bm}
\usepackage{latexsym}
\usepackage{amsfonts}
\usepackage{xcolor}
\usepackage{braket}
\usepackage{rotating}
\usepackage{array}
\usepackage{subcaption}  
\usepackage{float}
\usepackage{natbib}
\usepackage{cleveref}

\graphicspath{{./figs/}{./}}

\DeclareUnicodeCharacter{2212}{-}
\DeclareUnicodeCharacter{2061}{}

\def\br{\mathbf{r}}

\def\h2o{\mathrm{H}_2\mathrm{O}}

\date{\today}

\begin{document}

\title{Calculation and analysis of exciton couplings via a subsystem formulation of the $GW$-Bethe--Salpeter Equation}


\author{Sarathchandra Khandavilli}
\affiliation{Department of Chemistry and Pharmaceutical Sciences, Vrije Universiteit, De Boelelaan 1108, 1081 HZ Amsterdam, The Netherlands}

\author{Arno Förster}
\affiliation{Department of Chemistry and Pharmaceutical Sciences, Vrije Universiteit, De Boelelaan 1108, 1081 HZ Amsterdam, The Netherlands}

\author{Lucas Visscher}
\email{l.visscher@vu.nl}
\affiliation{Department of Chemistry and Pharmaceutical Sciences, Vrije Universiteit, De Boelelaan 1108, 1081 HZ Amsterdam, The Netherlands}

\begin{abstract}
We present a fragment-based framework for analyzing exciton couplings within the $GW$–BSE formalism using localized molecular orbitals, and assess how excitonic states in molecular dimers can be decomposed into local and charge-transfer (CT) sectors. Our localization procedure preserves orbital orthonormality via a block-diagonal unitary transformation, enabling a simple and interpretable analysis of excitonic interactions. Using ethylene and pyrene dimers as model systems, we identify key effects of excitonic basis truncation and coupling approximations on excitation energies. We then extend the method to chlorophyll dimers, where weak CT asymmetries emerge due to geometric distortions. This framework offers a tractable route to analyze excitonic behavior in complex systems and paves the way for future fragment-based reconstruction of full exciton coupling matrices in large molecular assemblies.
\end{abstract}


\maketitle

\section{Introduction}
\label{sec:intro}
One of the most critical issues in quantum chemistry is accurately describing excited states\cite{Loos2020f}, particularly charge-transfer (CT) excitations\cite{Loos2021, mester2022}, which play a fundamental role in processes such as photosynthesis\cite{Cheng2009,Curutchet2017} and photovoltaic energy conversion\cite{Gruber2012,Mikhnenko2015}. 
While time-dependent\cite{Runge1984, Petersilka1996} density-functional theory\cite{Hohenberg1964, Kohn1965} (TD-DFT) can be applied to large molecules\cite{Tamura2021, Sirohiwal2022, Volpert2023, Capone2023a}, it is often not accurate enough to provide definitive analyses and predictions. TD-DFT calculations tend to underestimate CT excitation energies, since the kernels of commonly used semilocal exchange-correlation functionals do not sufficiently describe the non-locality of CT excitations\cite{Tozer2003, Dreuw2004}. Range-separated\cite{leininger1997} hybrid functionals\cite{Yanai2004}, which incorporate non-local Hartree-Fock exchange at long range, help address this problem\cite{Chai2008, Rohrdanz2009}, though achieving good accuracy typically requires systematic tuning of the range-separation parameters\cite{stein2009reliable, Refaely-Abramson2011, Kronik2012} and careful adjustment of the short- and long-range exact exchange fractions, which are highly system-specific.\cite{Loos2021}

Computationally tractable alternatives to TD-DFT include equation-of-motion (EOM) coupled cluster (CC)\cite{Monkhorst1977, Stanton1993} and the $GW$\cite{Hedin1965, Golze2019} (where $G$ represents the single-particle Green's function and $W$ represents the screened Coulomb interaction)-Bethe-Salpeter equation (BSE) method\cite{Strinati1988, Onida2002, Blase2018}. Despite a canonical scaling of $N^6$, thanks to recent algorithmic developments\cite{Dutta2016, Dutta2018a} EOM-CC with single and double excitations (EOM-CCSD) can be applied to systems of the order of 100 atoms\cite{Sirohiwal2020}. Unfortunately, its accuracy diminishes for larger molecules, where it is even outperformed by the lower-scaling second-order CC (CC2) method\cite{Loos2020e, Loos2021a, Veril2021} and systematically overestimates CT excitation energies\cite{Knysh2024a}. Its similarity-transformed variant\cite{Nooijen1997,Nooijen1997a} performs somewhat better\cite{Loos2021} but requires the selection of an active space. $GW$-BSE, on the other hand, has a moderate canonical computational cost, and a low-scaling implementation has enabled applications to systems with thousands of correlated electrons\cite{Forster2022c}. $GW$-BSE accurately describes singlet excited states in weakly correlated systems\cite{Bruneval2013, Forster2022c, McKeon2022} also with Rydberg\cite{Gui2018a} or CT\cite{Gui2018a, Bhattacharya2024, Knysh2024a, Forster2025} character. $GW$-BSE is therefore a suitable method for the calculation of CT states in large molecular complexes of interest in photochemistry or photovoltaics. 

While the accuracy of the electronic structure method is important for a quantitatively and qualitatively correct description of excited states, the characterization of excited states poses an additional challenge. CT excitations are difficult to characterize due to the spatially delocalized nature of the excitonic wavefunction in the molecular orbital (MO) basis of large supramolecular systems. This spatial delocalization induces mixing between local and CT states, complicating their identification and analysis. Accurate description of these states is crucial, as they are precursors to charge separation in photosynthetic processes and to the development of photovoltaic technologies\cite{Mikhnenko2015, Gruber2012}.

Although CT states can be identified by visualizing molecular orbitals involved in the transition, such a characterization might sometimes fail and is not systematic. Hence, several systematic descriptors, based for instance on the overlap between the occupied and virtual states involved in the transition, orbital centroids, natural transition orbitals, or attachment/detachment densities have been proposed\cite{Peach2008, plasser2012, Guido2013, Plasser2014, Plasser2014a} (See Ref.~\citenum{Loos2021} for a recent review). While useful for characterizing CT excitations, the fact that they all rely on different criteria illustrates the non-trivial nature of characterizing CT states in the molecular orbital picture.

Partitioning the system into smaller subsystems\cite{Manby2012, Jacob2024} allows to analyze the character of excited states in a conceptually clear manner in terms of localized orbitals\cite{Neugebauer2007, johannes2019, Sen2022} and can reduce computational complexity\cite{Curutchet2017}.
Several proposed methods in this field can broadly be classified into two categories: bottom-up and top-down approaches. In bottom-up embedding schemes, the wave function of the total quantum many-body system is reconstructed from the wave functions of its subsystems or fragments. In this approach, inter-subsystem interactions are added to the subsystem Hamiltonians. In contrast, top-down embedding approaches start with the full Hamiltonian and wave function, but (initially) neglect correlations between specific fragments, simplifying the problem by focusing first on the most relevant interactions. In the context of DFT we refer the readers to References~\citenum{Tolle2022}~and~\citenum{bensberg2019automatic} for more details on these methods.

In subsystem TD-DFT\cite{Jacob2024} one may distinguish between approaches based on the orbital-free formalism of Weslowski and Warshel\cite{Wesolowski1993} which employs non-orthogonal orbitals and thus requires use of a kinetic energy density functional approximation, and the projection-based embedding method suggested by Manby and coworkers\cite{Manby2012} where orbital orthogonality between the subsystems is enforced. Both approaches allow for precise characterization of excited states in terms of local\cite{Neugebauer2007} and CT\cite{johannes2019,Sen2022} character. A versatile implementation of sub-system TD-DFT\cite{Neugebauer2009, Tolle2019, Tolle2022} was realised by the Neugebauer group\cite{serenity_pub}. They follow a bottom-up approach, initially partitioning the systems into smaller fragments and then constructing the supermolecular density from smaller fragment calculations. Alternatively, one may employ a top-down approach and achieve subsystem orthogonality by rotating supermolecular Kohn-Sham (KS) orbitals into fragment-localized orbitals\cite{Senjean2021} through Pipek-Mezey\cite{Pipek1989} localization. In both approaches, the locality of the resulting orbitals enables a straightforward definition of local and CT excitations, as well as analysis of their mixing within TD-DFT. 

Going beyond TD-DFT, several works have implemented the $GW$-BSE method in a projection-based embedding framework\cite{Baumeier2012, Wehner2017,Tolle2021, Tirimbo2023, Sundaram2024,Mauricio2024}. The formalisms developed in these works furthermore employ additional QM/MM embedding techniques\cite{Baumeier2014, Tirimbo2020, Tirimbo2020a, Duchemin2016, Li2016, Li2018} which allow to build effective excitonic models in condensed phases. 
In the current work, we introduce a novel $GW$-BSE embedding scheme based on fragment-localized molecular orbitals\cite{Sen2023}. In particular, we demonstrate that our top-down approach provides easy access to quasi-diabatic states and allows for the straightforward calculation of Coulombic and CT-mediated couplings. Such couplings between diabatic states are crucial for interpreting molecular spectra in molecular aggregates\cite{Hestand2015, Hestand2018} or analyizing photochemical processes\cite{Segatta2019, Popp2021}. While we are currently working on extending our approach to condensed phases via a QM/MM treatment, we here restrict ourselves to gas-phase calculations. In section~\ref{sec:theory} we briefly review our construction of molecular orbitals, following \citet{Sen2023} and show how this treatment generalizes to $GW$-BSE. In Section~\ref{sec:bench}, we give a detailed analysis of the excited states, obtained at different levels of $GW$ self-consistency and with TD-DFT, of a few model dimers at different intermolecular separations as well as a sample application to analyze CT interactions in a chlorophyll dimer. Section~\ref{sec:conclusion} concludes this work.

\section{Theory}

\label{sec:theory}

In this section, we summarise the use of localized orbitals in the linear response formalism of TD-DFT and $GW$-BSE.

\subsection{Optical excitations via linear response}

Within TD-DFT and $GW$-BSE, optical excitations are accessed by diagonalizing an effective two-particle Hamiltonian of the form
\begin{equation}
\label{excitonicHamiltonian}
    H = H_0 + K \;.
\end{equation}
Here, $H_0$ describes non-interacting electron-hole pairs, while the kernel $K$ describes their interactions.
Focusing on singlet excitations and using real-valued spin-restricted orbitals to define a basis of particle-hole states, Eq.~\eqref{excitonicHamiltonian} is typically expressed as (for derivations see for instance Refs.~\citenum{Sander2015, Bruneval2016a}),
\begin{equation}
\begin{pmatrix}
\mathbf{A} & \mathbf{B} \\
-\mathbf{B} & -\mathbf{A}
\end{pmatrix}
\begin{pmatrix}
\mathbf{X}_S \\
\mathbf{Y}_S
\end{pmatrix}
=
\begin{pmatrix}
\mathbf{X}_S \\
\mathbf{Y}_S
\end{pmatrix}
\Omega_S\;,
\end{equation}
with the matrix elements
\begin{equation}
\label{AandB}
A_{ia;jb} = \delta_{ij}\delta_{ab}\omega_{ia}  + K^{Hxc}_{ia;jb}
\end{equation}
and
\begin{equation}
B_{ia;jb} =   K^{Hxc}_{ia;bj} \;.
\end{equation}
$\omega_{ia;\sigma}$ is a single-particle energy difference, and $\Omega_S$ is the excitation energy for state $S$. A simplification often employed is the Tamm-Dancoff Approximation (TDA)\cite{Hirata1999}, which neglects the coupling matrix $B$, reducing the problem to the eigenvalue problem
\begin{equation}\label{TDA}
\mathbf{A}\mathbf{X} = \mathbf{X} \boldsymbol{\Omega} \;.
\end{equation}
In this work, we exclusively work in the TDA.

\subsubsection{Linear response with TD-DFT}

Within TD-DFT, the coupling matrix elements become
\begin{equation}
\label{KHxc}
\begin{aligned}
K^{Hxc}_{ia,jb} = & (ai|v|jb) + (1-c_x) (ai|f_{xc}|jb) - c_x (ab|v|ji)
\;,
\end{aligned}
\end{equation}
containing the four-center integrals given in Mulliken notation
\begin{equation}
\label{eq:eri}
( pq | v | rs ) = \int d\br \int d\br'
   \varphi^*_{p}(\br) \varphi_{q}(\br) v(\br,\br')
   \varphi^*_{r}(\br') \varphi_{s}(\br') \;,
\end{equation}
and the KS orbitals $\varphi$, and in Eq.~\eqref{AandB} $\omega_{ia} = \epsilon_{a} - \epsilon_{i}$. We follow the standard convention and let $p,q,r,s, \dots$ denote general, $i,j, \dots$ occupied, and $a,b, \dots$ virtual orbitals. $\epsilon$ denotes a KS eigenvalue. We allow for a $c_x$ amount of admixture of exact exchange besides the static xc-kernel $f_{xc}$, which is the functional derivative of the exchange-correlation potential  $V_{xc}$
\begin{equation}
\label{fxc_def}
    f_{xc}(\br,\br') = \frac{\delta V_{xc}(\br)}{\delta n(\br')} \;.
\end{equation}

\subsubsection{Linear response with $GW$-BSE}
In the BSE, the functional derivative \eqref{fxc_def} is replaced by the kernel
\begin{equation}
\label{Ivertex}
    K(3,5,4,6) = i \frac{\delta \Sigma(3,4)}{\delta G(6,5)} \;.
\end{equation}
Here, $\Sigma$ is the non-local, non-Hermitian, and frequency-dependent electronic self-energy, and $G$ is the single-particle Green's function. A number hereby denotes a combined space-time-spin index, $1 = (\br_1, t_1, \sigma_1)$. Within the $GW$ approximation\cite{Hedin1965} to the self-energy, $\Sigma$ is
\begin{equation}
\label{sigma}
    \Sigma(1,2) = iG(1,2) W(1,2) \;,
\end{equation}
where the screened Coulomb interaction $W$ is typically calculated within the RPA. 
Using \cref{sigma,Ivertex}, neglecting contributions of second order in $W$, and taking the static limit, one obtains the kernel matrix elements\cite{Rohlfing2000}
\begin{equation}
\label{K_BSE}
\begin{aligned}
K^{BSE}_{ia,jb} = &(ai|v|jb) - (ab|W(i\omega=0)|ji)\;.
\end{aligned}
\end{equation}
An important difference with TD-DFT furthermore arises from the fact that the dominant diagonal contribution $\omega_{ia}$ to the $A$ matrix in $H_0$ is constructed from the $GW$ QP energies $\epsilon^{QP}_{p}$ instead of the KS ones, $\omega_{ia} = \epsilon^{QP}_{a} - \epsilon^{QP}_{i}$. In the $G_0W_0$\cite{Hybertsen1985, Hybertsen1986, Rostgaard2010, Golze2019} and eigenvalue-only self-consistent $GW$ (ev$GW$)\cite{Blase2011}, the QP energy $\epsilon^{QP}_p$ is given by the solution of the expression
\begin{equation}
    \epsilon_{p} - \left[v_{xc} + \Sigma_{xc}\right]_{pp}(\omega) = \omega 
\end{equation}
with the largest QP weights.
In $G_0W_0$, only a single iteration of this scheme is performed, while ev$GW$ iteratively updates the QP energies. In our implementation\cite{Forster2021a} of the quasi-particle self-consistent $GW$ (qs$GW$) method
\cite{Faleev2004, VanSchilfgaarde2006, Kotani2007}, an effective, static, Hermitian QP Hamiltonian with matrix elements
\begin{equation}
\label{QPpart}
    \delta_{pq} \epsilon_q + \delta_{pq} \text{Re} \left[\Sigma (\epsilon_p)\right]_{pq} + (1-\delta_{pq}) \text{Re} \left[\Sigma (\omega=0)\right]_{pq} 
\end{equation}
is diagonalized instead (mode B in Ref.~\citenum{Kotani2007}), and it is understood that the self-energy contains the Hartree contribution. In eq.~\eqref{QPpart}, $\epsilon_p$ denotes the QP Hamiltonian in the basis that diagonalizes the previous iteration's QP Hamiltonian. The frequencies at which each matrix element of the self-energy is evaluated are not unique\cite{Kotani2007, Ismail-Beigi2017, Marie2023a}, but they typically do not affect the results much\cite{Forster2021a, Marie2023a}. For reasons of numerical stability, the construction of eq.~\eqref{QPpart} is typically preferred when the self-energy is calculated via analytical continuation as done in the current work\cite{Forster2021a, Lei2022, Harsha2024a}.

\subsection{Localized molecular orbitals}
We use the two-step procedure as outlined in references~\citenum{Senjean2021}~and~\citenum{Sen2023} for constructing localized fragment orbitals. The initial step involves constructing intrinsic fragment orbitals (IFOs), a generalization of Knizia's intrinsic atomic orbitals concept\cite{Knizia2013}. The first set of IFOs comprises the occupied and valence virtual orbitals. This set of orbitals has the same size as a minimal atomic orbital set for the system under consideration, but by construction spans the occupied MO space exactly. These orbitals are localized, but since they mix occupied and virtual orbitals, they can not be directly utilized in TD-DFT or $GW$-BSE. They are instead used as a basis-set-insensitive reference for a Pipek-Mezey\cite{Pipek1989} type localization in which the occupied orbitals are unitarily transformed into fragment-localized orbitals. The same procedure is followed for the valence virtual orbitals that complement this IFO space. In the second step, we follow Ref.~\citenum{Sen2023} and construct a second set of IFOs that comprises low-energy virtuals not contained in the valence virtual space. These already localized orbitals are in the current work used directly, instead of following an additional localization by rotating the low-energy virtuals as was done in Ref.~\citenum{Sen2023}. For convenience, we will briefly summarise the procedure currently implemented in ADF to show the differences with Ref.~\citenum{Sen2023} to which we refer for a more complete description of this procedure. An extensive description of the projection techniques used to form IFOs can be found in Ref.~\citenum{Senjean2021}. 

\begin{enumerate}
    \item Optionally remove core orbitals from the set of orbitals to be localized (to prevent mixing between core and valence-occupied orbitals during the localization).
    
    \item Define reference fragment valence orbitals by selecting a minimal MO space for each fragment. 
    
    \item Compute the valence IFOs. These IFOs fully span the occupied space, but span only part of the virtual space.
    
    \item Localize occupied orbitals and then recanonicalize within the fragment space to obtain valence occupied recanonicalized intrinsic localized molecular orbitals (RILMOs).
    
    \item Localize the valence virtual orbitals contained in the IFO space.
    \item Define orbitals spanning the virtual space complementary to the orbitals obtained in step 5.
    
    \item Re-canonicalise these complementary virtual orbitals to obtain semicanonical supermolecular virtual orbitals with effective energies $\epsilon^\prime$.
    
    \item Select orbitals with values of $\epsilon^\prime$ below a user-defined threshold.
    
    \item Define reference virtual fragment orbitals by selecting a suitable hard virtual space for each fragment.
    
    \item Compute the hard virtual intrinsic fragment orbitals.
    
    \item Recanonicalize the set of valence and hard virtual intrinsic fragment orbitals for each fragment to obtain the virtual RILMOs.
\end{enumerate}

Steps 1 through 10 are identical to the procedure described in Ref.~\citenum{Sen2023}. In that work, an additional localization step was performed in between steps 10 and 11 to prevent mixing-in of high-energy virtual orbitals. We found by further testing that this additional step did not improve results significantly and therefore simplified the algorithm by directly using the hard virtual IFOs in the final step.

In contrast to the GW-BSE-DIPRO scheme of \citet{Wehner2017}, wherein the eigenstates of an approximate Hamiltonian diabatic states are constructed from monomer electron-hole states, here diabatic states are obtained by first carrying out a supramolecular calculation and then localizing, thus retaining some interfragment interaction.  An important feature of our localization procedure is the preservation of the orthogonality of the orbitals. The final set of molecular orbitals is, like the original molecular orbitals, fully orthonormal and related to this set via a transformation matrix $\mathbf{U}$:
\begin{equation}
\label{Umatrix}
\phi^A_{i_A} = \sum_j U_{i_Aj}\phi_j;\ \phi^A_{a_A} = \sum_b U_{a_Ab}\phi_b \;.
\end{equation}
The matrix $U$ is rectangular, with the number of selected reference virtual orbitals determining the size of the virtual space in the local molecular orbital (LMO) basis. This truncates the virtual space relative to the canonical molecular orbital (CMO) basis, which in turn may be smaller than the space spanned by the atomic orbital basis in case of removal of nearly linear dependent functions.  

While the localization was initially applied with Hartree-Fock or KS orbitals, the procedure can equally well be applied to the $GW$ orbitals and their eigenvalues resulting from diagonalizing the qs$GW$ Hermitian QP Hamiltonian. In the G$_0$W$_0$ and ev$GW$ approaches the DFT orbitals remain unchanged and only the KS eigenvalues need to be replaced before the localization procedure.

\subsection{Linear response with localized orbitals}\label{lin_response_with_loc}

To characterize the excitonic states in terms of local and CT contributions, the linear response equation given in \cref{TDA}, originally defined in the supermolecular MO basis, is transformed to the fragment-specific orbital basis using the unitary transformation defined in \cref{Umatrix}, yielding for $GW$-BSE 
\begin{equation}
\begin{aligned}     
\label{ABSElocal}
    A_{i_Aa_C,j_Bb_D} =&  \delta_{a_C b_D}\Sigma^{QP}_{i_A j_B}-  \delta_{i_Aj_B} \Sigma^{QP}_{a_C b_D} \\
    &+ (a_Ci_A|v|j_Bb_D) \\
    &-(a_Cb_D|W(i\omega=0)|j_Bi_A) \;. 
\end{aligned}
\end{equation}
In the canonical QP basis, the QP Hamiltonian is diagonal, but this feature is only partly restored by the recanonicalization procedure, leaving off-diagonal elements between the different fragments that make \cref{ABSElocal} slightly more involved than \cref{AandB}. As an example, we consider two fragments, labeled A and B. In this basis, the linear response equation assumes the form 
\begin{widetext}
    \begin{equation}\label{subsystem_Amatrix}
\begin{pmatrix}
{\mathbf{A}}^{L_A} & {\mathbf{A}}^{L_A|L_B} & 
{\mathbf{A}}^{L_A|CT_{AB}} & {\mathbf{A}}^{L_A|CT_{BA}} \\
{\mathbf{A}}^{L_B|L_A} & {\mathbf{A}}^{L_B} & {\mathbf{A}}^{L_B|CT_{AB}} & {\mathbf{A}}^{L_B|CT_{BA}} \\
{\mathbf{A}}^{CT_{AB}|L_A} & {\mathbf{A}}^{CT_{AB}|L_B} & {\mathbf{A}}^{CT_{AB}} & {\mathbf{A}}^{CT_{AB}|CT_{BA}} \\
{\mathbf{A}}^{CT_{BA}|L_A} & {\mathbf{A}}^{CT_{BA}|L_B} & {\mathbf{A}}^{CT_{BA}|CT_{AB}} & {\mathbf{A}}^{CT_{BA}}
\end{pmatrix}
\begin{pmatrix}
{\mathbf{X}}^{L_A} \\
{\mathbf{X}}^{L_B} \\
{\mathbf{X}}^{CT_{AB}} \\
{\mathbf{X}}^{CT_{BA}}
\end{pmatrix}
=
\begin{pmatrix}
{\mathbf{X}}^{L_A} \\
{\mathbf{X}}^{L_B} \\
{\mathbf{X}}^{CT_{AB}} \\
{\mathbf{X}}^{CT_{BA}}
\end{pmatrix}
\boldsymbol{\Omega}
\end{equation}
\end{widetext}

where $L_A$ and $L_B$ denote local excitonic states in fragments A and B, respectively, while $CT_{AB}$ corresponds to an electron transfer from fragment A to B, and $CT_{BA}$ corresponds to an electron transfer from fragment B to A.
In this excitonic basis, we can diagonalize each of the diagonal blocks separately, giving quasi-diabatic (QD) states that are localized on either A or B or describe a specific CT between the fragments. 
\begin{equation}
\mathbf{A}^{QD} {\bar{\mathbf{X}}}^{QD} = \bar{\mathbf{X}}^{QD}\boldsymbol{\bar{\Omega}},
\end{equation}
with $\mathbf{A}^{QD} \in \{\mathbf{{A}}^{L_A},\mathbf{{A}}^{L_B},\mathbf{{A}}^{CT_{AB}},\mathbf{{A}}^{CT_{BA}}\}$.

In contrast to Rodríguez‑Mayorga et al.\cite{Mauricio2024}, who construct normalized but non‑orthogonal Frenkel‑exciton and CT trial states and solve a Sylvester's equation, our starting electron-hole states are already orthogonal. While this choice streamlines the treatment of the coupling matrices, it makes interpretation less straightforward as localized orthogonal orbitals necessarily have small tails at other subsystems. To study the coupling between the two lowest locally excited states, we compute
\begin{equation}\label{raw_coupling}
    \bar{\mathbf{J}}_{11}^{L_A|L_B}
    = \sum_{i_A,a_A,j_B,b_B}
      \bar{\mathbf{X}}^{QD}_{i_Aa_A,1}\,
      {\mathbf{A}}_{i_Aa_A,j_Bb_B}^{L_A|L_B}\,
      \bar{\mathbf{X}}^{QD}_{j_Bb_B,1}.
\end{equation}
Our method for obtaining the local-local couplings differs from that of \citet{Wehner2017}, who use first-order perturbation theory to incorporate CT states into local fragment excitations before calculating the couplings. In our work the coupling to CT states is explicit with orthogonal CT electron-hole states constructed from the localized orbitals. An approach similar to theirs, but not using perturbation theory, in which the full $\mathbf{A}$ matrix is partitioned according to the localized occupied orbitals, allowing excitations from these to all virtuals, is also implemented. This yields QD states that are partially corrected for CT, including electron but not hole transfer to other fragments, and can also form a basis for analysis. The corresponding coupling elements are then only labeled by the occupied space ($O_A$ or $O_B$) from which the excitations originate 
\begin{equation}\label{comp_coupling}
\bar{\mathbf{J}}_{11}^{O_A|O_B}\\
    = \sum_{i_A,j_B,a,b}
      \bar{\mathbf{X}}^{QD}_{i_Aa_A,1}\,
      {\mathbf{A}}_{i_Aa,j_Bb}^{O_A|O_B}\,
      \bar{\mathbf{X}}^{QD}_{j_Bb,1}.    
\end{equation}
We will showcase this possibility when analysing the pyrene dimer that was also studied by \citet{Wehner2017}.

\subsection{Technical Details}

\subsubsection{Computational Setup}
All calculations were performed using the Amsterdam density functional (ADF) engine of a locally modified version (version 2024.207) of the Amsterdam modeling suite (AMS)\cite{Baerends2025}. We used the all-electron triple-zeta polarised (TZP) basis set\cite{vanLenthe2003} in most calculations, but also performed some calculations using the aug-QZ6P basis set\cite{Forster2021} to assess basis set incompleteness errors. The LRC-$\omega$PBEh functional\cite{Rohrdanz2009} as implemented in LIBXC\cite{Lehtola2018} was used in all TD-DFT calculations and as starting point for all $GW$ calculations. We set the numerical quality to \textit{VeryGood} in all calculations. Most importantly, this includes the use of an auxiliary basis consisting of auxiliary functions with angular momentum up to $ l=6$ \cite {Forster2020} for the pair atomic density fitting (PADF)\cite{Spadetto2023} used in all DFT and $GW$-BSE calculations\cite{Forster2020b, Forster2022c}. In the aug-QZ6P calculations, we increased the size of the auxiliary basis by setting its quality to \textit{Excellent}. We further eliminate almost linear dependent products of basis functions from the primary basis by setting the $K$-matrix regularisation parameter to $5 \times 10^{-3}$\cite{Spadetto2023}.

\subsubsection{$GW$ calculations}

ADF implements the $GW$ method within a variant of the space-time method\cite{H.N.Rojas1995, Rieger1999, Liu2016, Wilhelm2018} which has been described elsewhere\cite{Forster2020b,Forster2021a}. In each iteration of a $GW$ calculation, this method requires performing three Fourier transforms between imaginary time and imaginary frequency domains on non-uniform grids\cite{Kaltak2014}. In particular, $W$ is calculated in imaginary frequencies and transformed to imaginary time to calculate $\Sigma$. We noticed that this step induces small instabilities in the calculations for certain geometries, leading to noisy potential energy surfaces. This issue is not caused by the choice of imaginary frequency or time grids, as they occur both with the minimax grids\cite{Kaltak2014} implemented in AMS\cite{Forster2021} and those implemented in the GreenX library\cite{Azizi2023,Azizi2024}. 

For this reason, we perform all $GW$ calculations in this work purely in the imaginary frequency domain, followed by analytical continuation to real frequencies.  This algorithm is, for instance, described in ref.~\citenum{Ren2012}. Using density fitting, these calculations scale quartic with system size, which is fast enough to calculate all the pyrene and pyridine dimers' potential energy surfaces (PES) on modest hardware. We use 32 modified Gauss-Legendre grid points in all calculations, chosen according to the prescription of ref.~\citenum{Fauser2021}. Only for the chlorophyll dimer, where we are not interested in calculating a PES, we employ the space-time algorithm using 20-point minimax grids native to AMS\cite{Forster2021}. 

When the self-energy is calculated on the imaginary frequency axis, each matrix element of the self-energy is calculated individually, and the computational effort of the calculation increases with the number of computed matrix elements. While in qs$GW$, all matrix elements are needed, in $G_0W_0$ or ev$GW$, one often only calculates a small subset of diagonal elements around the Fermi level. Then either a Hedin shift\cite{Pollehn1998} or a constant scissor-like shift is applied to the remaining QP energies (either for the subsequent BSE calculations or for the next ev$GW$ iteration)\cite{Vlcek2018b, Holzer2019, Wilhelm2021, Li2022a}. These shifts lead to noisy potential energy surfaces, and therefore, we always explicitly calculate the full QP spectrum.

\subsubsection{Localization procedure}
Unless otherwise indicated, default parameters for selecting hard virtuals were employed, i.e. a threshold of 2.0 Hartree for the supermolecular virtual orbital selection (step 8) and a threshold of 10.0 Hartree for the reference hard virtuals (step 9). Reference valence orbital spaces (step 2) were chosen according to the built-in definition of valences that follows standard chemical conventions (i.e. 1s2s2p for carbon and nitrogen, 1s for hydrogen). Core orbitals were not removed (step 1). The ADF implementation of TD-DFT and $GW$-BSE with LMOs allows for running a calculation in the LMO basis without applying restrictions on the excitations. These supermolecular results (which we will label LMO) can be directly compared to the  canonical molecular orbital (CMO) results to check the effect of the truncation of the virtual space in the transformation to the LMO basis. In all calculations, we partitioned the systems into two fragments, resulting in the coupling matrix comprising the blocks defined in \eqref{subsystem_Amatrix}. To keep the number of coupling elements to analyze manageable, we construct our QD excitonic basis using 10 LE states on each fragment and also 10 CT states between each fragment. With this total dimension of 40, it is trivial to diagonalize the full matrix \cref{subsystem_Amatrix} and obtain all its eigensolutions.
\section{Results and Discussion} 
\label{sec:bench}
For reference, we perform TD-DFT and $GW$-BSE calculations in the CMO basis. This procedure yields excitations of the dimer with a mixed character that can be directly compared to the coupled states resulting from the subsystem approach. As the kernel used in both approaches is identical, deviations between the two sets of results arise from truncations performed in the subsystem method. The first of these truncations is due to the selection of a subset of the full virtual space from each fragment in step 9 of the localization procedure. As we use a rather high threshold of 10 Hartree, this should not impact the target low-energy states much as can be verified by comparing the supermolecular LMO and CMO results. The second truncation is due to not solving and coupling all states of the local and CT subspaces. The impact of this truncation is small at larger intermolecular distances, where the adiabatic states are well-described as a linear combination of a small number of quasi-diabatic states, but becomes important at short distances, where also higher-energy quasi-diabatic states start to contribute.

\subsection{The ethylene dimer}
As the smallest model system to study $\pi\rightarrow\pi^*$ transitions, the ethylene molecule has been extensively studied by both experimentalists and theoreticians. As discussing all aspects of this model system would go beyond the scope of the present work, we mention only a few key references. Concerning the monomer, we refer to Feller, Peterson and Davidson\cite{feller2014ethylene} for a discussion of the lowest vertical transitions that highlights in particular the $^1B_{1u}$ or "V" state (using the labels for the excited states of Mulliken\cite{mulliken1979}) arising from the $\pi\rightarrow\pi^*$ transition. A key question\cite{lindh1989theoretical} is the diffuseness of this V state, as this is quite dependent on the treatment of electron correlation. As the V state has its minimum at a twisted $D_{2d}$ geometry, it is not the lowest excited state at the ground state geometry; this is the Rydberg-like $^1B_{3u}$ "2R" state. Early EOM-CCSD calculations\cite{WattsBartlett1996} gave benchmark values of 7.28 and 7.98 eV for these vertical 2R and V transitions. The most accurate values from full configuration interaction quantum Monte Carlo (FCIQMC)\cite{daday2012full-d50} and selected configuration interaction (sCI)\cite{Loos2018b} place the V state in the range 7.89 to 7.96 eV, depending on the choice of equilibrium geometry. Looking at results for the dimer and larger oligomers, we note the extensive study by Michl and coworkers\cite{zaykov2019singlet}, who considered coupling between the diabatic states considered in the current work as well as to the state arising from singlet coupling of two monomer triplet states. While we can in our implementation also compute localized triplet states, we have not yet considered generalizing our treatment to include these excitations. Results for only the Frenkel excitonic and charge separated states as a function of dimer distance and amount of exact exchange in the density functional approximation are presented by Norman and Linares\cite{norman2014interplay} and show a large splitting of the lowest excitonic states at a separation distance of 3.5 $\AA$. Using the symmetry of a perfectly stacked system they assign the lowest, dipole-forbidden, $^1B_{3g}$ state of the dimer to a Frenkel exciton and assign the first dipole-allowed $^1B_{2u}$ state to its coupling partner. 

Turning now to our calculations, we first consider a distance of 5 \AA, a regime of weak exciton coupling in which it is easy to relate the dimer states to the monomer states. In Table~\ref{tab:ethylene} we list the obtained monomer, QD, coupled and supermolecular energies.

\begin{table*}[htb]
\label{tab:ethylene}
\caption{
Calculated vertical excitation energies (eV)for ethylene dimers 5\,\AA, listing local ($L_A$), CT, coupled (CP) and supermolecular values in the LMO and CMO bases. Monomeric (M) excitation energies for the 2R and V states and are listed as well.}
\label{tab:ethylene}
\begin{ruledtabular}
\begin{tabular}
{lcccccccccc}
Method &
$\Omega^{L_A}_{2R}$ & $\Omega^{L_A}_{V}$ &
$\Omega^{CT_{AB}}_{S_1}$ & $\Omega^{CP}_{S_1}$ &
$\Omega^{LMO}_{2R}$ & $\Omega^{LMO}_{V}$ &
$\Omega^{CMO}_{2R}$ & $\Omega^{CMO}_{V}$ &
$\Omega^M_{2R}$ & $\Omega^M_{V}$ \\

\midrule
TD-DFT   & 7.685 & 8.487 & 8.738 & 7.541 & 7.528 & 8.539 & 7.529 &  8.539  & 7.626 & 8.480 \\
$G_0W_0$ & 7.483 & 8.432 & 10.158 & 7.364 & 7.354 & 8.577 & 7.356 & 8.585 & 7.425 & 8.514 \\
ev$GW$   & 7.529 & 8.490 & 10.218 & 7.409 & 7.400 & 8.634 & 7.399 & 8.634  & 7.466 & 8.571 \\
qs$GW$   & 7.768 & 8.589 & 10.330 & 7.648 & 7.640 & 8.754 & 7.658 & 8.740 & 7.742 & 8.555 \\
\end{tabular}
\end{ruledtabular}
\end{table*}

Looking first at the monomer data, we note a $\sim$0.5 eV difference between our highest level of theory, qs$GW$, and the EOM-CCSD values\cite{WattsBartlett1996} for the 2R and V states. This difference for the 2R state can be attributed to the basis set incompleteness\cite{WOOD1974} of the TZP basis. While usually sufficiently accurate for medium to large-sized molecules, TZP does not contain diffuse enough functions to describe ethylene's Rydberg excitations. When the same calculation is performed in the more diffuse aug-QZ6P basis\cite{Forster2021}, the lowest monomer energy becomes $7.40$ eV using the qs$GW$ method. This is a deviation of $0.12$ eV to the 2R states in the EOM-CCSD reference and a deviation of $0.04$ eV to the sCI reference from the QUEST database\cite{Loos2018b}. Improving the accuracy of the V state is more difficult, in the aug-QZ6P basis energy calculated with its qs$GW$-BSE is 8.44 eV, only a minor improvement relative to the TZP basis. Without the TDA, we however obtain an excitation energy of 8.01 eV, in excellent agreement with accurate reference values\cite{daday2012full-d50, Loos2018b}. We refer to Angeli\cite{Celestino2009} for an in-depth discussion of the different aspects that should be considered to get a quantitatively correct excitation energy. While computed too high relative to the low-lying Rydberg states within the TDA, the V state is, however, easy to identify due to the symmetry of the monomer as well as by its dominant $\pi$ to $\pi^*$ character.

At a distance of 5\AA\  we see that the local excitation energies $\Omega^{L_A}_{2R}$ remain rather close to the monomer value, with the interaction with the other ethylene slightly increasing the excitation energy of the $2R$ state. For the V state the main effect is that the distortion of the symmetry (originally $^1B_{2u}$ in the monomer) due to the presence of the other monomer allows for mixing with an almost degenerate $^1B_{2g}$ Rydberg state. This gives rise to two local states, the lower one being primarily the V state, but with some Rydberg character, the upper one the original $^1B_{2g}$ Rydberg state with some valence character. At this intermolecular distance, CT states are for the $GW$ variants at too high energy to contribute, so the lowest coupled state (S$_1$) consists primarily of the local 2R states. The state with the highest oscillator strength is S$_{10}$ and is composed of the local V states with some admixture of Rydberg states that was already present in the uncoupled localized states. Also for this state, we do not see any appreciable CT character. Looking specifically at the CT states before coupling, we also clearly see the limitations of density functional approximations to describe CT as the differences between TD-DFT and all $GW$-BSE variants is 0.2 eV or less for the local QD states, but more than 1 eV for the CT state.   

\begin{figure}[hbt!]
\centering
   \begin{subfigure}[c]{0.45\textwidth}
    \centering
        \includegraphics[width=\textwidth]{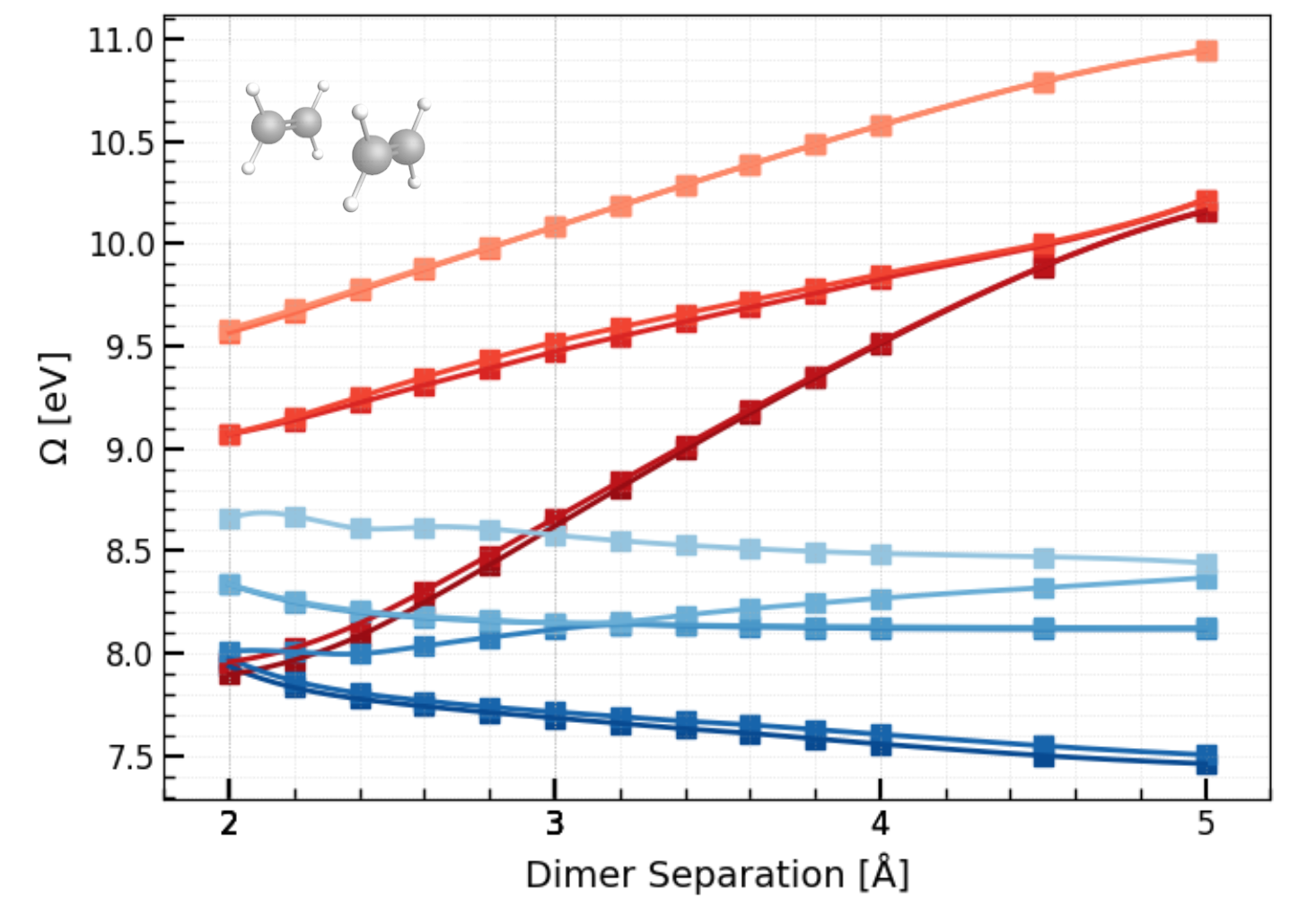}
    \end{subfigure}
    \begin{subfigure}[c]{0.47\textwidth}
    \centering
        \includegraphics[width=\textwidth]{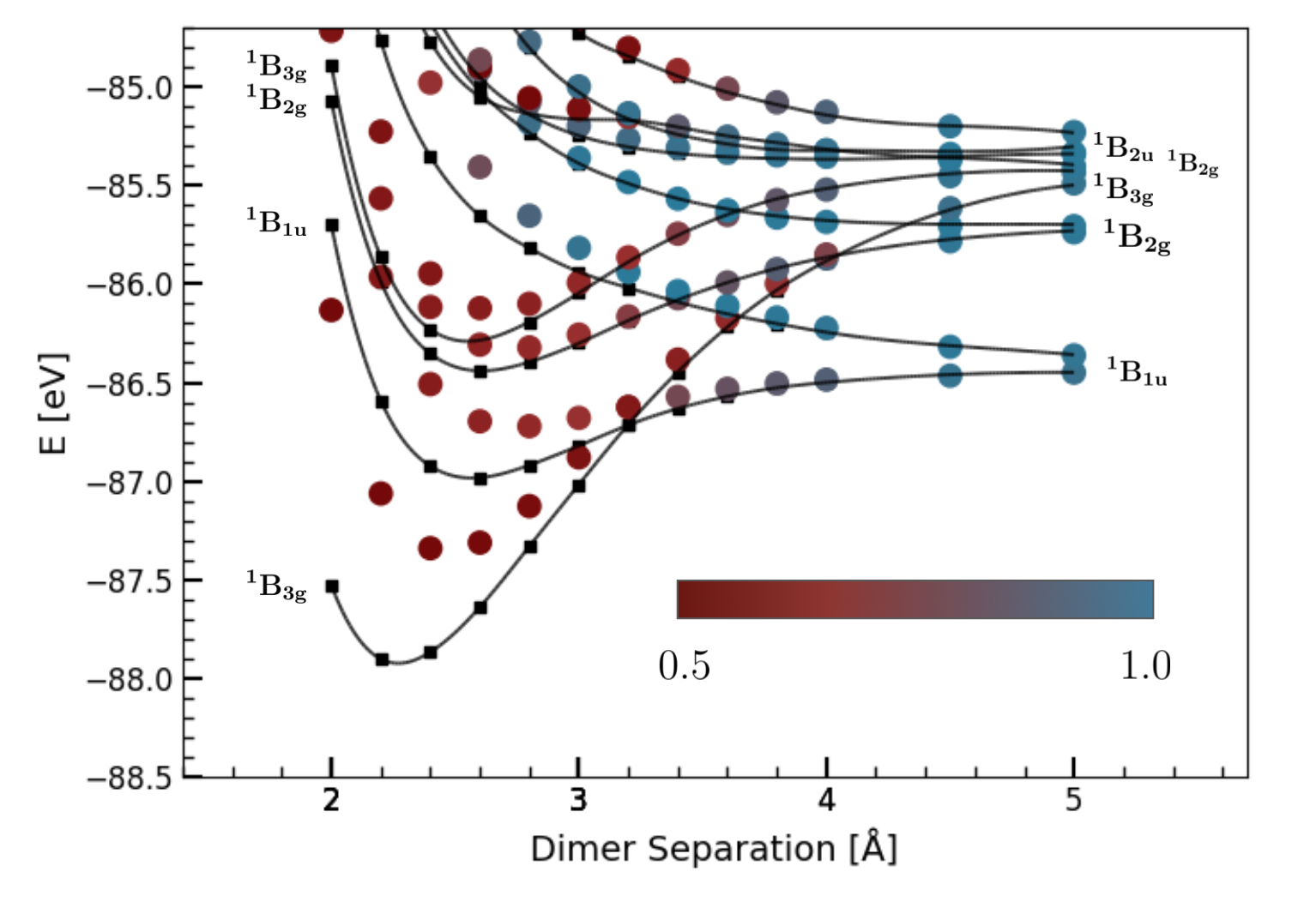}
    \end{subfigure}
     \hfill
 \caption{Upper panel:  Excitation energies computed using $G_0W_0$ within the QD basis, obtained by diagonalising the local (shades of blue) and charge-transfer (CT) excitation blocks (shades of red) separately.
 Lower panel: Dimer bonding energies (ground state) plus excitation energies computed using $G_0W_0$-BSE. Discrete blue/red markers correspond to the coupled states in the QD basis, black squares represent singlet excited states calculated in the CMO basis and the continuous black line is the interpolation of the black squares. Symmetry labels on the left denote the supramolecular (dimer) symmetry of the CMO states, whereas labels on the right indicate the underlying monomer fragment symmetries.
}
\label{fig:coupled_ethylene}
\end{figure}

The picture changes if we move the ethylene molecules closer together. At 3 \AA\ the lowest excited state is now of $^1$B$_{3g}$ symmetry and is composed for 38\% of the symmetric combination of the original V states. As shown in Figure \ref{fig:coupled_ethylene}, there is very significant admixture of CT character at this intermolecular distance. This also evident in the $^1$B$_{1u}$ state that is again found as the 10$^{th}$ excited state and consists for 38\% of the antisymmetric combination of the original V states. Also for this state, we now see significant CT character. The states arising from the local Rydberg states lie in between these two valence states and show varying amounts of admixture of CT states. As can be seen by the proximity of the QD excitation energies to the CMO values, the CMO result can in the region 3 to 5 $\AA$ be approached well by coupling only a limited number of QD states. At distances shorter than 3 $\AA$ the agreement deteriorates due to the incompleteness of the QD basis. Such errors are common to all subsystem approaches, see for instance the discussion for projection-based $GW$–BSE embedding schemes in reference~\citenum{Sundaram2024}. They can be reduced by calculating and coupling more QD states. Computing and coupling all QD basis states should yield excitation energies close to those in the CMO basis, with the only error remaining coming from the initial truncation of the virtual orbital space made when computing the localized orbital space. To verify that convergence is still possible, also when using the QD expansion rather than full diagonalization, we carried out additional calculations by increasing the number of CT states at a dimer separation of 2.2~\AA, which is close to the minimum for the lowest excitonic state in this constrained sandwich geometry. At this distance we see strong interaction leading to a lowest singlet excitation energy ($S_1$) in the CMO basis of 3.00~eV, while in the LMO basis it is 2.99~eV. This indicates that deviation between the QD and CMO result does, as expected, not arise from the slight truncation applied when localizing the orbitals. With the 10 CT states used in our original calculations, the excitation energy in the QD basis is 3.84~eV, a deviation of 0.84~eV. When the number of CT states is increased to 40, the QD excitation energy decreases to 3.67~eV, and further increasing the number of CT states to 80 decreases this further to 3.18~eV, a still significant deviation but indicating convergence as well as the importance of CT states at short distances.

To disentangle the effects of the different couplings on the adiabatic state energies, we plot in Figure \ref{fig:coupled_ethylene} the energies obtained by coupling either only the LE QD states or only the different CT states. Looking at these curves, we see for the latter no visible splittings, which is expected as couplings between these sectors depend on overlaps between orbitals localized at different subsystems. Couplings will not be numerically zero as the choice of working with orthonormal localized orbitals necessarily leads to a small tail of each localized orbital on the other system. This leads to a small dipole-dipole interaction that will vanish at long distance. For the LE states we see an appreciable splitting for especially the pair of states resulting from the monomer V states. The splitting shows the expected Förster type decay of $R^{-3}$ with intermolecular distance $R$. For the CT states, the decomposition shows the characteristic $R^{-1}$ dependence that puts them below the LE states at short distance. Visible are furthermore the Pauli repulsion effects that push the Rydberg orbitals and associated excitations to higher energies as the monomers approach each other. This clear picture that emerges illustrates the usefulness of our definition of local orbitals, either as a convenient basis for a full diagonalization or, in the weak coupling regime, as a means to define quasidiabatic local states and analyze their couplings.

\subsection{Pyrene dimers}
Pyrene, a prototypical polycyclic aromatic hydrocarbon, has interesting properties, both as a monomer and as a dimer, the latter being a prototypical example of an exciplex\cite{Goldschmidt1970,Yoshihara1971}. 
For the monomer, the two lowest singlet excitations, $S_{1}\,(^{1}B_{2u})$ and $S_{2}\,(^{1}B_{1u})$, attracted early attention\cite{Geldof1969} because pyrene shows weak fluoresce from $S_{2}$, contrary to Kasha's rule\cite{Kasha1950}. 
This non-Kasha emission arises due to the large oscillator strength of $S_{2}$ and the small energetic spacing of these two states, leading to significant intensity borrowing from non-adiabatic mixing of the two states\cite{Majumdar2025}. An alternative view\cite{Braun2022} is that of a dynamic equilibrium between the two states requiring a sufficiently high vibrational temperature and a collision-free environment. Given the interest in this molecule, many calculations have been performed at various levels of theory\cite{Kerkines2009, Roos2018, Shirai2019, docasal2020, Braun2022, Majumdar2025}. As typically vertical excitation energies are reported, it is, also for the well-resolved\cite{Micheal1986} S$_0\rightarrow \rm{S}_1$ transition with a band origin of 3.37 eV, common to compare to band maxima\cite{Braun2022, Majumdar2025}. Our TD-DFT results show correct state ordering, in contrast to some previous TD-DFT results\cite{docasal2020}, but in qualitative agreement with the TDA (PBE0) results of Majumdar \textit{et al.}\cite{Majumdar2025}.

\begin{figure}[htb!]
\centering
    \begin{subfigure}[c]{0.47\textwidth}
    \centering
        \includegraphics[width=\textwidth]{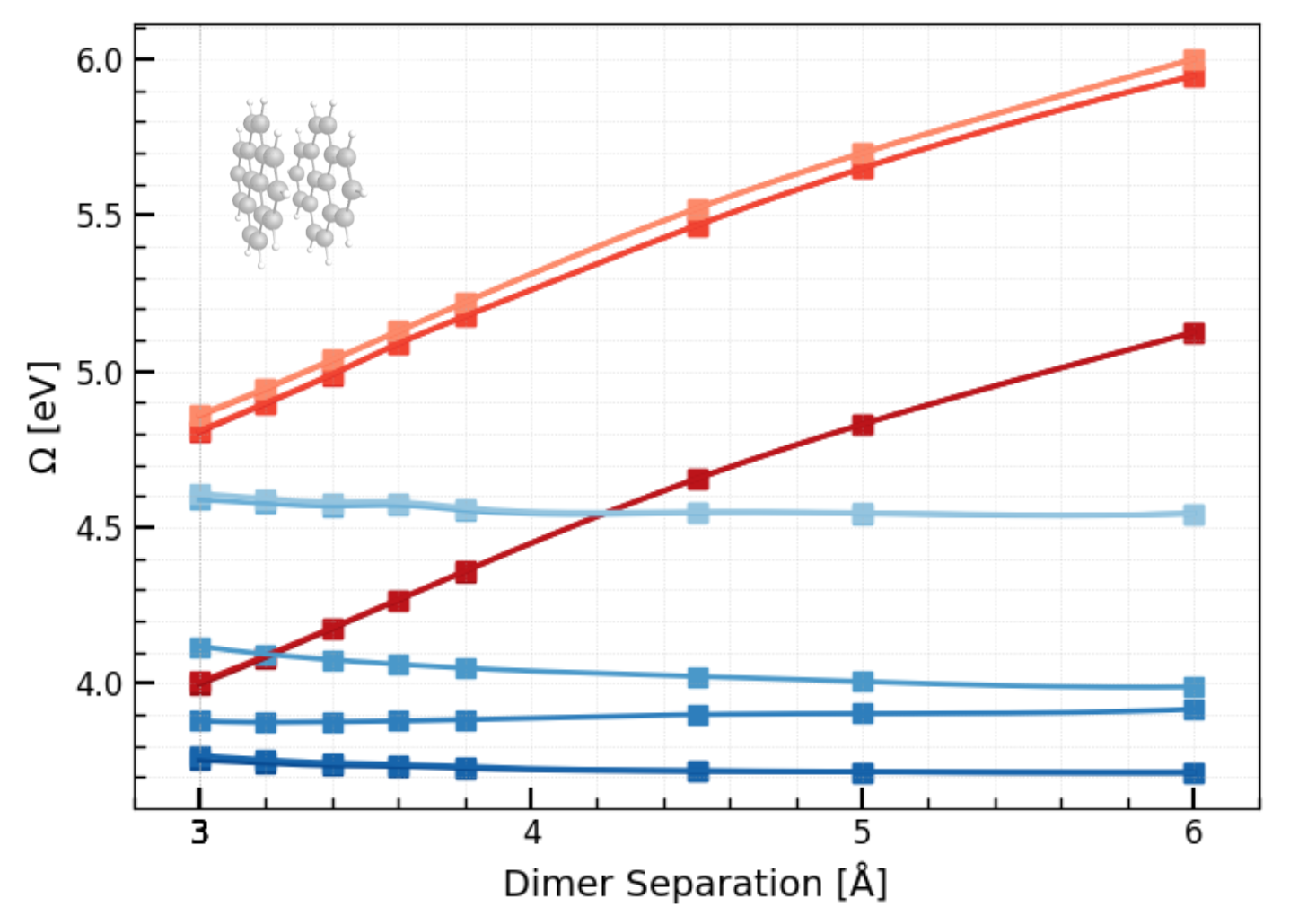}
    \end{subfigure}
    \begin{subfigure}[c]{0.47\textwidth}
    \centering
        \includegraphics[width=\textwidth]{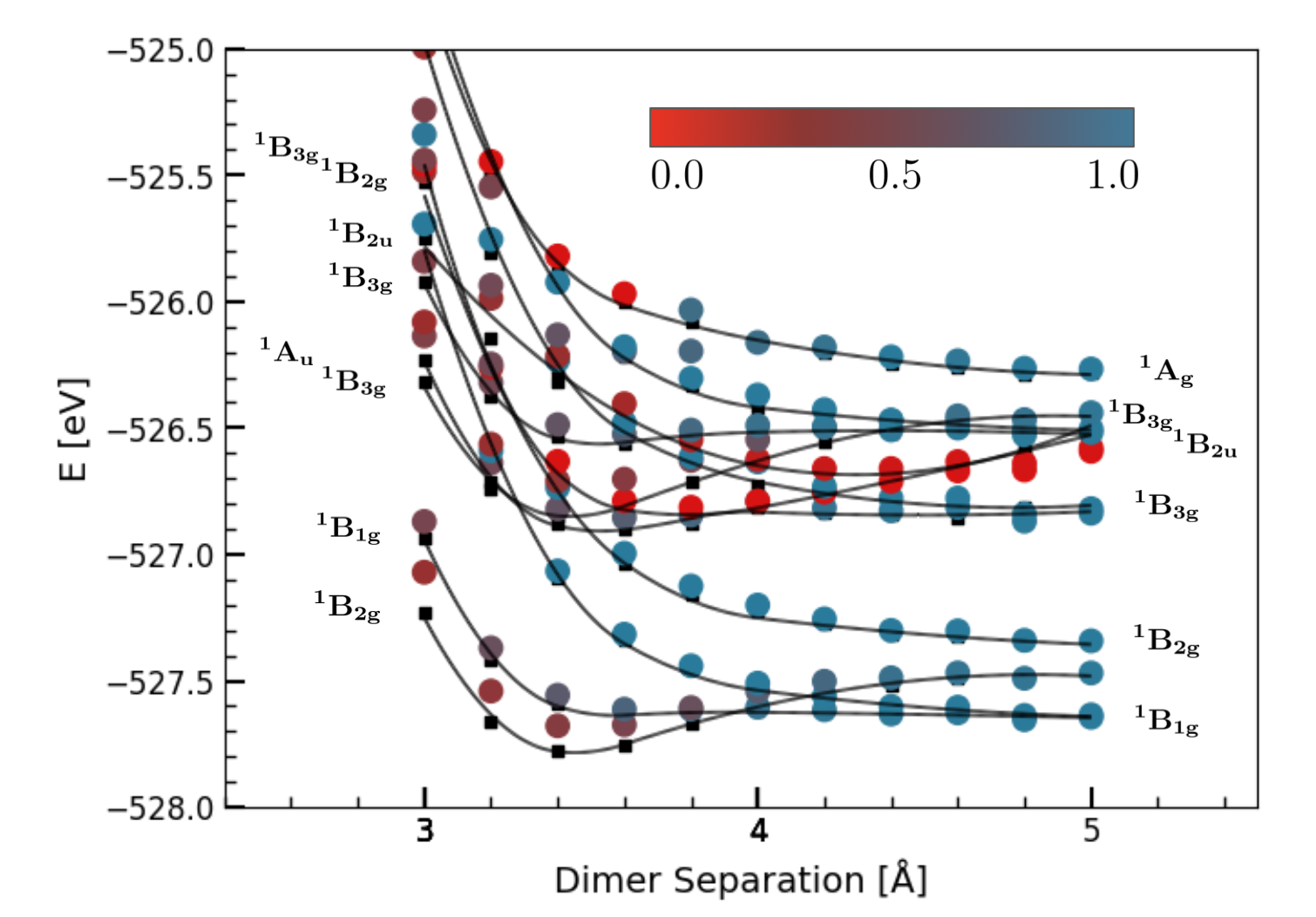}
    \end{subfigure}
     \hfill
 \caption{Upper panel:  Excitation energies computed using $G_0W_0$ within the QD basis, obtained by diagonalising the local and charge-transfer (CT) excitation blocks separately.
 Lower panel:  Lower panel: Dimer bonding energies (ground state) plus excitation energies computed using $G_0W_0$-BSE. Discrete blue/red/black (black indicating purely CT state, red indicating mixture and blue indicating purely a local excited state) markers correspond to the coupled states in the QD basis, black squares represent singlet excited states added to the bonding energies calculated in the CMO basis and the continuous black line is the interpolation of the black squares. Symmetry labels on the left denote the supramolecular (dimer) symmetry of the CMO states, whereas labels on the right indicate the underlying monomer fragment symmetries.}
\label{fig:coupled_pyrene}
\end{figure}

\begin{table*}[htb]
\caption{Vertical excitation energies of pyrene and their energy differences
$\Delta(S_2-S_1)$ (eV) at different levels of theory compared with band maxima
values\cite{Micheal1986, Ohta1987}.}
\label{tab:pyrene_monomer_vs_exp}
\begin{ruledtabular}
\begin{tabular}{lccccccc}
State &
TD-DFT &
$G_{0}W_{0}$ &
ev$GW$ &
qs$GW$ &
CC2\cite{Kerkines2009} &
MRX(4)\cite{Shirai2019} &
Exp. \\
{$S_{1}\,(^{1}B_{2u})$} & 3.981 & 3.711 & 3.700 & 3.750 & 3.675 & 3.40 & 3.46 \\
{$S_{2}\,(^{1}B_{1u})$} & 4.165 & 3.956 & 3.945 & 3.971 & 3.955 & 3.84 & 3.87 \\ $\Delta(S_2-S_1)$
& 0.184 & 0.245 & 0.245 & 0.221 & 0.280 & 0.44 & 0.41
\\
\end{tabular}
\end{ruledtabular}
\end{table*}

From Table \ref{tab:pyrene_monomer_vs_exp}, we observe that all methods corresponding to $GW$-BSE improve upon TD-DFT when judging the deviation of the $L_a$ and $L_b$ excitation energies with respect to the experimental band maxima. The results at our highest level of theory, qs$GW$-BSE are in reasonable agreement with the CC2 results of \citet{Kerkines2009}, while the MRX(4) variant of quasi-degenerate perturbation theory\cite{Shirai2019} gives the best agreement with experiment, both in terms of excitation energies and for the gap between them. For all single-reference methods, the gap between the two lowest excited states is clearly underestimated. This underestimation reflects the multi-reference character of the $^{1}B_{2u}$ state whose excitation energy Shirai \textit{et al.}\cite{Shirai2019} found to be much more dependent on the choice of reference than the $^{1}B_{1u}$ state.

Like in the ethylene case, we are again primarily interested in the excimer complex\cite{Winnik1993} that can be formed in the dimer. With the qualitatively correct state ordering found in the monomer, we like to study how the coupling between the $S_1$ and $S_2$ states (also called $L_b$ and $L_a$, respectively, in Platt's notation\cite{Platt1949}) is affected by the admixture of CT states. While this analysis is more easily done in a constrained perfect sandwich geometry, we first look at the minimum configuration, a slipped sandwich geometry reported by \citet{Lishcka2019}. We see in Table \ref{ref_comparison_pyrene} qualitative agreement of the LMO results with the NEVPT2 calculations of \citet{Lishcka2019}. Both methods indicate that the first 3 states have negligible oscillator strengths, while the fourth carries has a similar magnitude oscillator strength as found for the $S_{2}$ state of the monomer. The main difference is the much smaller splitting observed in our qs$GW$-BSE calculations, which is only 0.02 eV. Analysis of the coupled QD states reveals that this small splitting is caused by mixing of the upper component of the $S_2$ combination with CT states, which lowers the energy of that state relative to the lower component. Switching off the coupling with CT states and only allowing coupling of LE states increases this gap to 0.13 eV, with all four states approximately 0.08 eV above the NEVPT2 values. The sensitivity of the splitting to the inclusion of CT states provides a possible explanation for the spread of results for this pair of states reported in the literature. With DFT/MRCI\cite{Lishcka2019}, the bright state is, for instance, found as the third excited state, whereas in SOS-ADC(2)\cite{docasal2020} it is reported as the fourth state, 0.5 eV above the third excited state.

\newcolumntype{E}{>{\centering\arraybackslash}m{2.0cm}}

\begin{table*}[htb]
\small
\centering
\caption{qs$GW$ vertical excitation energies (eV) and oscillator strengths (in parentheses) for the pyrene dimers separated by 3.35\,Å in the slipped configuration taken from Ref.~\citenum{Lishcka2019}. LE-CP results are obtained by coupling only LE states, CP results include coupling to the CT states.}
\label{ref_comparison_pyrene}
\begin{ruledtabular}
\begin{tabular}{E E E E c E E}
\toprule
$\Omega^{LE}_{S_n}$ &  $\Omega^{CT}_{S_n}$ & 
$\Omega^{LE-CP}_{S_n}$ & $\Omega^{CP}_{S_n}$ & $\Omega^{LMO}_{S_n}$ & NEVPT2\cite{Lishcka2019}   \\
\midrule
3.855 (0.00) & 4.324 (0.03) & 3.854 (0.00) & 3.817 (0.00) & 3.749 (0.00) & 3.78 (0.00)  \\
3.855 (0.00) & 4.325 (0.03) & 3.856 (0.00) & 3.832 (0.00) & 3.760 (0.00) & 3.78 (0.00)  \\
4.079 (0.33) &  5.092 (0.00)           &   4.005 (0.00)  & 3.966 (0.00) & 3.870 (0.00) & 3.91 (0.00)   \\
 4.079 (0.33) &  5.092 (0.00)           &   4.134 (0.52)             & 3.977 (0.50) & 3.894 (0.40) & 4.05 (0.52)   \\
\bottomrule
\end{tabular}
\end{ruledtabular}

\end{table*}

 We now turn to discuss perfectly stacked monomers in which the interaction strength can be easily tuned to reach the strong interaction limit in which more involvement of CT states is expected. For computational convience we have computed interaction curves with the $G_0W_0$ approach which gives similar results as qs$GW$ (see for a comparison the $G_0W_0$ results for the slipped configuration in the Supporting Information) but at a significantly reduced computational cost in the MO basis. 

As shown in Figure \ref{fig:coupled_pyrene}, the Davydov splitting of the coupled $^1B_{2u}$ state is due to its large dipole transition dipole of 4.75 D already visible at an intermolecular distance of 5\AA. Admixture of CT states becomes important below 5 \AA\ with the QD 4$^{th}$ LE and lowest CT states crossing at about 3.2 \AA\
To display the CT character of the coupled QD state, we have colored the data points accordingly and we clearly see the influence of CT mixing as the monomers approach each other further. First for the states arising from the $^1B_{3g}$ monomer states at around 4.5 \AA\ , then at about 3.8 \AA\ also for the upper component of the coupled $^1B_{2u}$ states.  Since the CT state crosses the lowest LE states, the lowest state at the excimer equilbrium geometry is best characterized as a mixture of the original $^1B_{2u}$, $^1B_{1u}$ and a few CT states. As seen also for ethylene at shorter distances, the limited expansion in 10 QD states of each sector gives rise to some deviations between the CP and LMO results. As the qualitative picture is the same, however, an analysis in terms of QD states is still meaningful.

\begin{figure}[htb!]
\centering
    \begin{subfigure}[c]{0.47\textwidth}
    \centering
        \includegraphics[width=\textwidth]{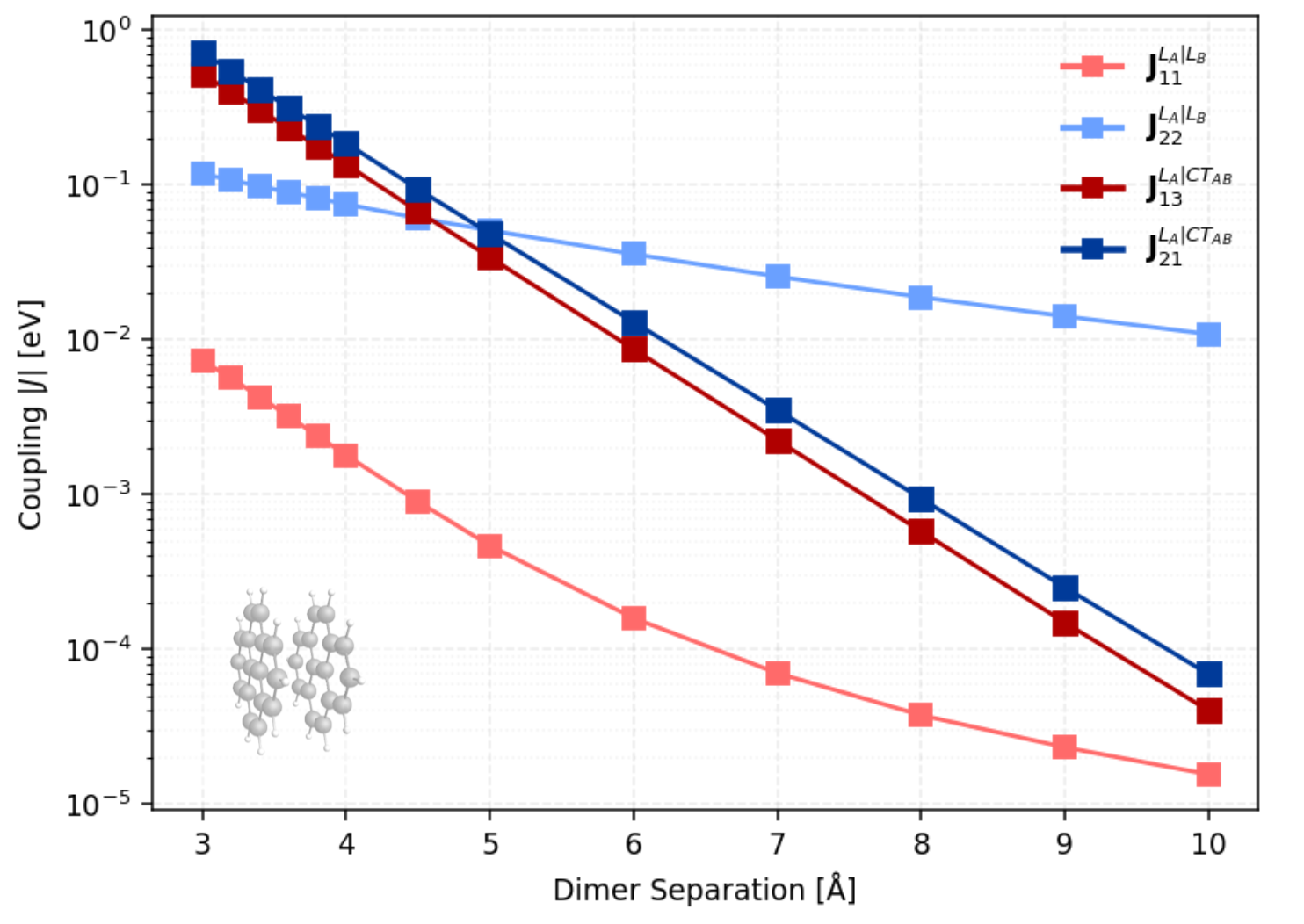}
    \end{subfigure}
    \begin{subfigure}[c]{0.47\textwidth}
    \centering
        \includegraphics[width=\textwidth]{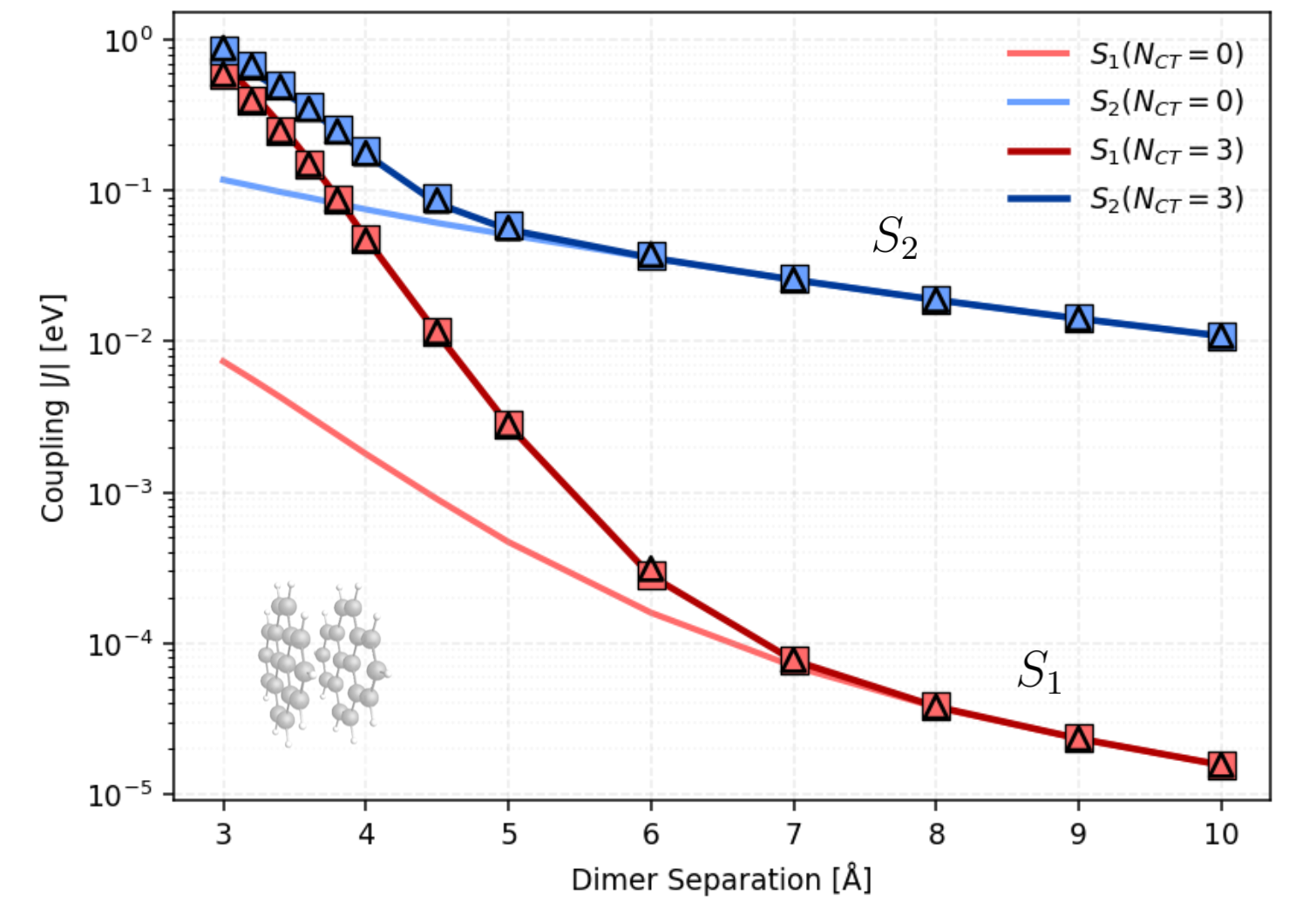}
    \end{subfigure}
     \hfill
 \caption{Upper panel : Largest couplings between QD states. Lines with shades of red show couplings of the first LE state, lines with shades of blue show couplings of the second LE state.
 Lower panel : Couplings between the first two LE states of the pyrene dimer. The blue and red square scatter points show the coupling in the CMO basis, computed as half the splitting of $B_{1g}$ and $B_{2g}$ states. The black triangular points represent couplings calculated by including all CT states via Eq. \ref{comp_coupling}. The lines (without markers) indicate couplings obtained by limiting the number of CT states included in the QD basis.}
\label{fig:davydov_splitting_s1_s2}
\end{figure}

Couplings between QD states are also interesting in their own right, as they can be used to define a tight-binding Hamiltonian for the study of exciton transport in molecular materials\cite{Mikhnenko2015, Popp2021}. To compare our computed coupling to other approaches to derive such couplings, in particular the $GW$-BSE-DIPRO method of Wehner and Baumeier\cite{Wehner2017}, we plot in the upper panel of Figure \ref{fig:davydov_splitting_s1_s2} the distance dependence of the most important couplings between QD states that we computed. We see the expected exponential decay with distance of the LE-CT coupling matrix element, but note that the coupling to the first CT state is at distances shorter than 5 \AA \ for $S_2$ even larger in magnitude than the slower decaying LE-LE coupling. For the weaker coupled $S_1$ states, the LE-CT coupling with the third CT state is larger even at 10 \AA, but because this state is then too high in energy to contribute, one can still use the F\"orster dipole approximation to compute the Davydov splitting between the states. This is displayed in the lower panel of \ref{fig:davydov_splitting_s1_s2} in which the Davydov coupling is displayed, computed from the splittings in the $B_{1g}-B_{2u}$ ($S_1$) and $B_{2g}-B_{1u}$ ($S_2$) pairs of dimer states. We see that at intermolecular distances below 7 \AA\ for $S_1$ or below 5 \AA\ one needs to include CT states to get accurate couplings. This corroborates the findings of \citet{Wehner2017}, but we note that, due to the use of orbitals from a supermolecular calculation that already describes polarization effects on the orbitals, we need to only include the three lowest CT states to get perfect agreement with the CMO reference calculations. An efficient way to obtain these couplings with CT-corrected local states directly is to only employ localized orbitals for the occupied orbitals and define QD states in which all excitations from these localized orbitals are taken into account. Results of these calculations are also displayed in the lower panel of Figure \ref{fig:davydov_splitting_s1_s2}. As lowest states of only two QD blocks needed to be computed, these calculations were about 2 times faster than with explicit determination of the two types of CT states in addition to the LE states.

\subsection{Chlorophyll dimers}
While the highly symmetric ethylene and pyrene molecules are convenient for detailed analysis and validation of our procedure, the construction of QD states and their coupling elements is typically desired for larger, less symmetric molecules. To show that the approach is easily applicable for this kind of systems, we will now consider a chlorophyll dimer system. We again construct the localized fragment orbitals and construct the lowest QD excitonic states with TD-DFT and different $GW$-BSE methods.

\begin{figure*}[hbt]
    \centering
    \includegraphics[width=0.8\linewidth]{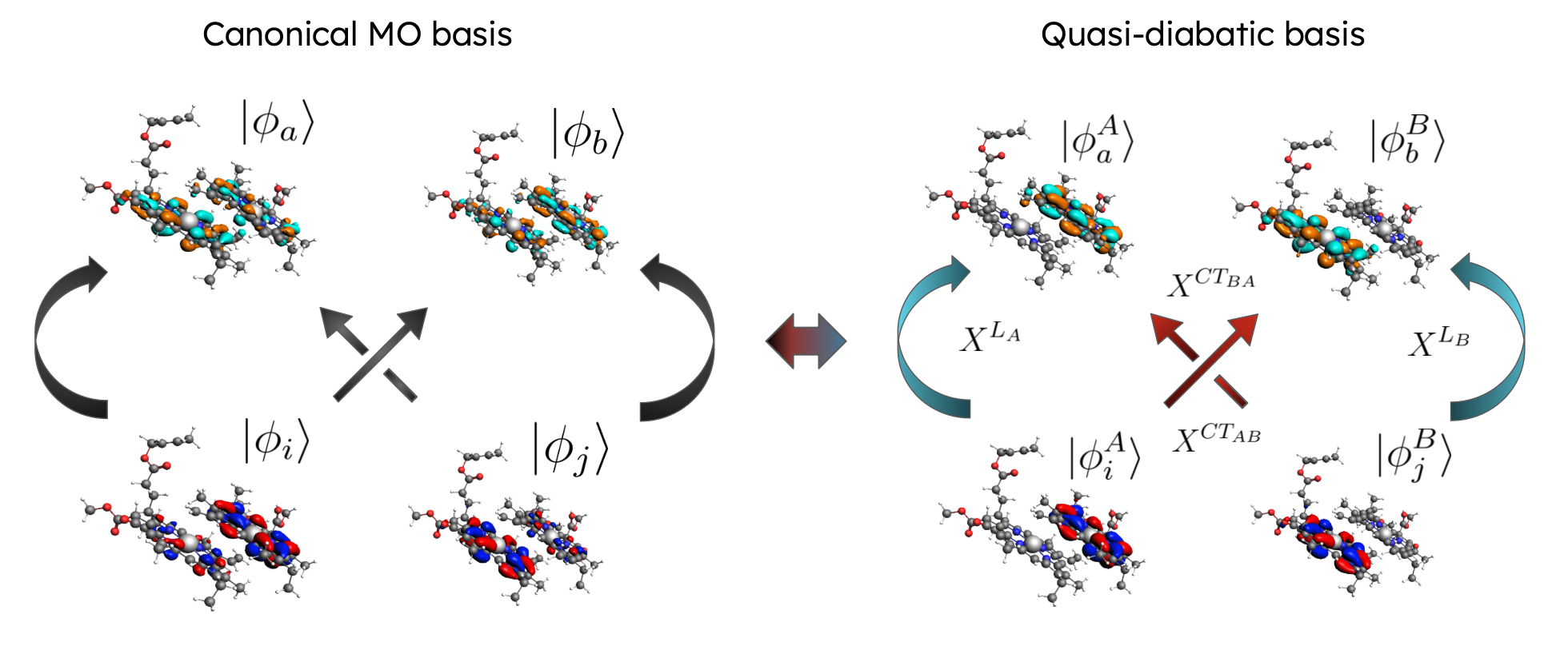}
    \caption{Comparison of canonical and localized molecular orbitals involved in excitation processes. Left: Canonical occupied \(|\phi_i\rangle\;|\phi_j\rangle\) and virtual \(|\phi_a\rangle\;|\phi_b\rangle\) orbitals of the full dimer system. Right: Corresponding localized orbitals on fragments A and B, showing occupied \(|\phi_i^A\rangle\), \(|\phi_j^B\rangle\) and virtual \(|\phi_a^A\rangle\), \(|\phi_b^B\rangle\) orbitals. The localized representation captures spatial separation, enabling fragment-based excitation analysis.}
    \label{fig:orbital_comparison}
\end{figure*}

\begin{table*}[htb]
\centering
\caption{
Calculated vertical excitation energies (eV) of the first singlet state ($S_1$) of the chlorophyll dimer, listing local excited states ($L_A; L_B$), their difference ($\Delta \Omega^{L}_{S_1}$), charge-transfer (CT) states and their difference ($\Delta \Omega^{CT}_{S_1}$), coupled (CP) QD states, CMO results ($\Omega^{CMO}_{S1}$), and deviations ($\Delta \Omega_{S_1}$) of the coupled QD to the CMO results.}

\label{tab:chl_dimer}
\begin{ruledtabular}
    
\begin{tabular}{
    l@{\hskip 15pt}  
    c@{\hskip 8pt}c@{\hskip 10pt}c@{\hskip 20pt}
    c@{\hskip 8pt}c@{\hskip 10pt}c@{\hskip 20pt}  
    c@{\hskip 8pt}c@{\hskip 10pt}c@{\hskip 10pt}
}
\toprule
\multirow{2}{*}{\textbf{Method}}  
 & \multicolumn{3}{c}{\textbf{Local excitations}}  
 & \multicolumn{3}{c}{\textbf{CT excitations}}  
 & \multicolumn{3}{c}{\textbf{}} \\
\cmidrule(lr){2-4} \cmidrule(lr){5-7} \cmidrule(lr){8-10}
 & $\Omega^{L_A}_{S_1}$ & $\Omega^{L_B}_{S_1}$ & $\Delta \Omega^{L}_{S_1}$ 
 & $\Omega^{CT_{AB}}_{S_1}$ & $\Omega^{CT_{BA}}_{S_1}$ & $\Delta \Omega^{CT}_{S_1}$ 
 & $\Omega^{CP}_{S_1}$ & $\Omega^{CMO}_{S_1}$ & $\Delta \Omega_{S_1}$ \\
\midrule
$G_0W_0$  &  2.206 & 2.201 &  0.005 &   2.656 &  2.516 &  0.140 &  2.104 &  2.027 & 0.077 \\
ev$GW$  & 2.175 & 2.170 & 0.005 & 2.619 & 2.479 & 0.140 & 2.076 &  1.998 & 0.078\\
qs$GW$  & 2.172 & 2.169 & 0.003 & 2.580 & 2.438 & 0.141 & 2.070 & 1.981 & 0.088\\
TD-DFT & 2.358 & 2.353 & 0.005 & 2.914 & 2.764 & 0.150  & 2.262 &  2.195 & 0.067\\
\bottomrule
\end{tabular}
\end{ruledtabular}
\end{table*}

As shown in Table~\ref{tab:chl_dimer}, TD-DFT, in this case, systematically overestimates both local and CT excitations relative to the $GW$-based methods. All methods show minimal differences in local excitation energies (also called "site energies" in this context) for the two fragments, with their difference ($\Delta\Omega_{S_1}^{L}$) consistently below 0.01 eV, as expected from the very similar structures of the two fragments. While site energies are very close, an asymmetry in the CT QD energies arises from the slight rotational offset in the placement of the two monomers. These CT excitation energies are given by \(\Omega_{S_1}^{CT_{AB}}\) and \(\Omega_{S_1}^{CT_{BA}}\), corresponding to CT from fragment \(A\) to \(B\) and vice versa. This asymmetry, \(\Delta\Omega_{S_1}^{CT}\) of 0.14 eV with $GW$-BSE (0.15 eV with TD-DFT), indicates a weak directionality in the CT states, which is usually attributed to the protein environment\cite{Tamura2021, Sirohiwal2023}.

The coupled excitation energy, ($\Omega_{S_1}$), is obtained by coupling all computed LE and CT states, while \(\Omega_{S_1}^{\text{CMO}}\) refers to the corresponding excitation energy in the canonical molecular orbital (CMO) basis. The discrepancy ($\Delta\Omega = |\Omega_{S_1} - \Omega_{S_1}^{\text{CMO}}|$) again reflects the limitations of the truncated QD basis in spanning the full excitonic space, with errors ranging from 0.067 eV (TD-DFT) to 0.088 eV (qs$GW$-BSE). To place the qs$GW$-BSE results in the context of existing literature, we compare them with the CMO excitation energy reported in our previous work\cite{Forster2022c}, where we obtained a value of 1.94~eV. In the present work, the corresponding excitation energy is slightly higher at 1.98~eV. This difference arises because Ref.~\citenum{Forster2022c} did not use the TDA.

Analyzing the couplings in more detail clarifies that CT states are crucial to get correct state couplings and Davydov splittings. The coupling between the $S_1$ states of the monomers is only 0.85 meV and increases to 32 meV when CT states are included via equation (\ref{comp_coupling}). The contribution of CT states is also visible when looking at the state composition, where we find for $G_0W_0$ the lower component of the Davydov split pair to consist of 52\% $L_B$, 30\% $L_A$, 12\% $CT_{BA}$ and 5\% $CT_{AB}$. The upper component is for 60 \% localized at monomer A and has with only 4\% significantly less CT character. The asymmetry between the energies and contributions of \( CT_{AB} \) and \( CT_{BA} \) to the lowest coupled state is interesting, because such asymmetries can indicate a preferred directionality in exciton transport. More study, in particular including a model for environment effects of the surrounding protein scaffold that influences site energies and couplings is, however, needed\cite{Sirohiwal2020a, Sirohiwal2022, Sirohiwal2023, Tamura2021}. For such future applications use of equation (\ref{comp_coupling}) is probably most efficient, with the explicit determination of QD CT states serving as a tool for detailed analysis.

\section{\label{sec:conclusion}Conclusion}

We introduced a fragment-orbital-based framework for analyzing exciton couplings within the $GW$–BSE formalism, built on a block-diagonal unitary transformation that preserves the orthonormality of the molecular orbitals. Unlike density-based embedding schemes such as frozen-density embedding\cite{Tolle2021}, which aim to reduce computational cost by embedding a subsystem into a larger environment, our approach is not primarily intended as a cost-saving scheme. Instead, it enables a physically transparent analysis of excitation processes in terms of localized fragment states. In this regard, our work is conceptually closer to projection-based methods\cite{Sundaram2024,Mauricio2024, Wehner2017}, which also aim to partition excitations across subsystems, though typically without preserving strict orthonormality between fragment orbitals.

Our approach has two characteristic features. The first is the use of intrinsic fragment orbitals which resemble the original fragment orbitals but are polarized in the supramolecular potential. As these orbitals form an orthogonal set, we obtain a simple computational scheme that can be used with various electronic structure methods. The second feature is the description of CT effects, which can either be analyzed in detail by explicit evaluation of couplings between QD states, or be folded in by constructing QD states that consist of coupled local and electron transfer excitons.

Using ethylene and pyrene dimers as a minimal model, we first validated our localized excitonic analysis framework by recovering expected trends in local and CT excitations across different separations and $GW$-BSE variants. These initial results establish how basis truncation and coupling approximations influence excitation energies. Building on this, we applied our method to chlorophyll dimers—systems of biological relevance—revealing near-degenerate local excitations and small but consistent asymmetries in CT states due to geometric asymmetry. In the model dimer that we studied, we find local–CT couplings to be sizable and essential for a proper description of exciton coupling. This example demonstrates that localized orbital analysis within $GW$-BSE not only simplifies interpretation but also offers a tractable path for distinguishing exciton transport in complex molecular systems. 

Vibronically dressed electronic excitation in chromophoric complexes are described by diagonalizing the combined Frenkel-Holstien Hamiltonian \cite{Spano2024,Hestand2018}. This requires excited state gradients, as electronic excitations are coupled to different vibronic modes \cite{Bogdain2025,Nagami2020}. Computing such gradients within the $GW$-BSE framework remains an active area of research \cite{Bogdain2025,Alrahamneh2025}. Our procedure supplies the requisite site energies $\epsilon_m$ and $J_{mn}$ which are needed for its construction and thereby provides a good starting point for further investigations.

\section*{Supplementary Material}
In the supplementary material, we have added results of additional calculations that support the findings of this article as well as the employed structures of the molecules.

\section*{Author Declarations}

\subsection*{Conflict of Interest}
The authors have no conflicts to disclose.
\subsection*{Author Contributions}
S. K., A. F. and L.V. contributed to this work in different ways. A Contribution Role Taxonomy (CRediT) statement is given below in accordance with the NISO standard.\\

\textbf{Sarathchandra Khandavilli}: Data curation; Formal analysis(equal); Investigation; Methodology (supporting); Project administration (supporting); Resources (supporting); Software(lead); Validation; Visualization; Writing – original draft(equal); Writing – review \& editing (supporting).
\\
\textbf{Arno Förster}: Methodology (supporting); Supervision(equal); Writing – original draft(equal); Writing – review \& editing (supporting).
\\
\textbf{Lucas Visscher}: Conceptualization; Formal analysis(equal); Funding acquisition; Methodology (lead); Project administration (lead); Resources (lead); Software(supporting); Supervision(equal); Writing – review \& editing (lead).

\section*{Acknowledgments}
We gratefully acknowledge insightful discussions with Erik van Lenthe, Souloke Sen, and Peiting Xue. SK and LV acknowledge funding via the project Divide \& Quantum (project number 1389.20.241) of the research programme NWA which is (partly) financed by the Dutch Research Council (NWO). AF acknowledges funding through a VENI grant from NWO under grant agreement VI.Veni.232.013. Use of the supercomputer facilities at SURFsara was sponsored by NWO Physical Sciences, with financial support from the Netherlands Organization for Scientific Research (NWO).

\bibliography{export_p, all}

\end{document}